\documentclass{emulateapj}

\slugcomment{Draft version August 9, 2004}
\shortauthors{Rafelski \& Zaritsky}
\shorttitle{SMC Star Cluster Age Distribution}

\begin{document}
\title{The Star Clusters of the Small Magellanic Cloud: \\ Age Distribution}

\author{Marc Rafelski and Dennis Zaritsky}
\email{marcar@astro.ucla.edu, dzaritsky@as.arizona.edu}

\affil{Steward Observatory, University of Arizona, 933
   North Cherry Avenue, Tucson, AZ 85721, USA}

\begin{abstract} 
We present age measurements for 195
star clusters in the Small Magellanic Cloud based on comparison of integrated colors measured from the Magellanic Clouds Photometric
Survey with models of simple stellar populations. We find that
the modeled nonuniform changes of cluster colors with age can lead to spurious 
age peaks in the cluster age distribution, that
the observed numbers of clusters with age, $t$, declines smoothly as $t^{-2.1}$, 
that for an assumed initial cluster mass function scaling as $M^{-2}$ the 
dependence of the cluster disruption time on mass is proportional to $M^{0.48}$,  that
despite the apparent abundance of young clusters the dominant epoch of cluster formation was the initial one, and that there are significant differences in the spatial distribution
of clusters of different ages. 
Because of limited precision in our age measurements,
we cannot address the question of detailed correspondence between the cluster
age function and the field star formation history. However, this sample provides an
initial guide for which clusters to target in more detailed studies of specific age intervals.
\end{abstract} 


\section{Introduction}

The discovery of young star clusters in ongoing galaxy mergers \citep{holtzman92, whitmore} has
provided added motivation for understanding the formation mechanism
of stellar clusters and the dynamical processes that lead to their destruction.
While the ancient clusters in  our own galaxy provide only  indirect information regarding the
formation or disruption of clusters, the Magellanic Clouds
contain numerous young clusters ($< 1$ Gyr old) for study \citep{hodge61, van81}. 
The large number of clusters is of great statistical value, but it has divided the
studies into two camps. Ages have been determined
either for large samples of clusters using 
low-precision measurements, such as integrated colors 
(for example see \cite{van81}) or for small samples using high-precision
measurements, such as color-magnitude diagrams (for 
an early example see \cite{baird}). Each approach has its strengths and weaknesses,
but a study of the overall population must include age estimates for a large fraction
of all the clusters in the galaxy. Only with such measurements can one begin to 
address whether the cluster formation history tracks the field star formation history,
how quickly clusters are dissolved, and whether the cluster system as a whole
still retains some memory of cluster formation episodes.

The study of the clusters in the Magellanic Clouds has a  rich history that cannot
be properly summarized here. Nevertheless, from that work we know
of several interesting features in the age distribution of clusters in both the Large
and Small Magellanic Clouds (LMC and SMC). 
In the LMC there is an ``age gap" (a deficit of clusters of ages
corresponding to ages between roughly 3 and 10 Gyr) that has been extensively 
studied \citep{jensen, dacosta, van91, rich01}. 
The age distribution of clusters in the SMC is less  established.
Ground based studies of SMC clusters have concluded that the
age distribution is continuous  \citep{dh98, msf98}, but analysis of  
the SMC's seven brightest old (age $>1$~Gyr) clusters with
the {\it Hubble Space Telescope} has suggested two distinct 
episodes of cluster formation that occurred 2 $\pm$ 0.5 Gyr and $8\pm2$~Gyr 
ago \citep{rich00}. Three additional SMC clusters that satisfy the age criteria but 
are too faint to be included in the $HST$
study are Lindsay~1, which has a ground-based age of 9~Gyr and so is coincident
with the older burst, Lindsay 11, which has an age of 3.7 Gyr,
and Lindsay 113, which has an age of 5.3 Gyr \citep{msf98, rich00}. 
Neither of the latter two appear 
to have formed within the two identified bursts and may populate the ``gap".
Whether the structure in the age distribution is real, and whether it tracks any
pattern observed in the field star formation history of the SMC, are open questions.

We present a study based on integrated colors of 195 clusters 
in the SMC that incorporates a
number of improvements over similar previous studies: 1) we use
digital imaging in four filters (most previous studies of integrated colors 
were based on photographic plates and utilized only two or three filters), 2) we use
the structural parameters individually derived for each cluster \citep{hill03} to set our aperture size
(most previous studies have a fixed aperture size for all clusters), and 3) 
we use multiple theoretical models of simple stellar populations to derive an 
age and associated uncertainty (most previous studies calibrated integrated
colors to a small number of clusters with ages derived from color-magnitude
diagrams --- a procedure that can lead to coarse age resolution and 
unknown systematic problems that are difficult to explore, see \S3.1).
The data and methodology used to measure ages are
described in \S2. We discuss various aspects of the derived age distribution in \S3
and summarize in \S4.

\section{Data and Analysis}

\subsection{Images of the Cluster Sample}

\begin{figure}
\plotone{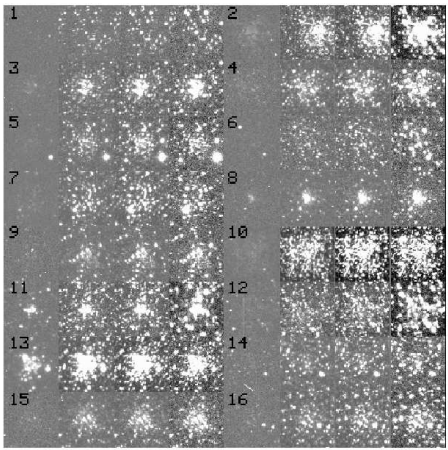}
\caption{$UBVI$ images of the clusters. Each row contains the four filter images for two clusters.
The images are 70 arcsec on a side and oriented such that N is to the right and E to the bottom.
This Figure includes the images for clusters 1 through 16. The remainder of the images are
available in the electronic version.
\label{mosaic}}
\end{figure}

The sample of SMC clusters that we analyze is presented by \cite{hill03}. Coordinates
and cross-identifications are given in Table \ref{tab1}. With the exception
of clusters and associations embedded in the complex environment of star forming
regions, these 204 clusters are the most unambiguous clusters in the 4.5$^\circ \times 4^\circ$
section of the SMC observed by the Magellanic Clouds Photometric Survey \citep{zar02}. Figure \ref{mosaic} shows images of clusters 1 through 16.
We have excluded
clusters and associations that lie within emission line nebulae because of the
difficulty in measuring integrated colors in such an environment. 
The absence of these clusters explains in part why other catalogs list many more
star clusters (for example see \cite{ogle} and \cite{bica00}), but we also found from
visual inspection of over a thousand candidate clusters that many are quite marginal. 
It is not our contention that only the clusters presented here are real.
There are almost certainly real clusters 
lurking among the previously published catalogs that we have excluded.
However,  even if real, we cannot confidently measure the 
colors or structural parameters of such poor clusters.
When considering the results from
our sample, one must remain aware that we are incomplete in both the youngest
and poorest (poor in either total number of stars or surface density) clusters. 
The latter might be a particularly interesting population because it
may consist of clusters that are nearing the end of their lifetime as bound objects.

\begin{deluxetable*}{crrrrrrrrrrr}
\tabletypesize{\scriptsize}
\tablecaption{CLUSTERS AND THEIR COLORS 
\label{tab1}}
\tablewidth{0pt}
\tablehead{
\colhead{Number} & 
\colhead{RA}   & 
\colhead{DEC}   &
\colhead{$M_{V} $} &
\colhead{$\sigma_{V}$} &
\colhead{U$-$B} & 
\colhead{$\sigma_{U-B}$} &
\colhead{B$-$V}  & 
\colhead{$\sigma_{B-V}$} &
\colhead{V$-$I}  &
\colhead{$\sigma_{V-I}$} &
\colhead{Name\tablenotemark{a}}
}

\startdata

1 &0 24 18.67 &$-$73 59 35.8 &15.92 &1.17 &0.75 &0.24 &0.72 &0.14 &0.92 &0.07 &\nodata \\
2 &0 24 43.16 &$-$73 45 11.7 &\nodata &\nodata &\nodata &\nodata &\nodata &\nodata &\nodata &\nodata &K5, L7, ESO28-18 \\
3 &0 25 26.60 &$-$74 $\,$ 4 29.7 &14.23 &\nodata &0.57 &\nodata &0.65 &\nodata &1.10 &\nodata &K6, L9, ESO28-20 \\
4 &0 27 45.17 &$-$72 46 52.5 &13.95 &0.38 &0.03 &0.09 &0.80 &0.04 &1.20 &0.05 &K7, L11, ESO28-22 \\
5 &0 28  $\,$ 2.14 &$-$73 18 13.6 &14.85 &1.17 &0.28 &0.05 &0.61 &0.05 &1.00 &0.40 &K8, L12 \\
6 &0 29 55.22 &$-$73 41 57.1 &14.31 &1.12 &-0.05 &0.48 &0.65 &0.06 &0.37 &0.32 &HW3 \\
7 &0 30  $\,$ 0.26 &$-$73 22 40.7 &13.84 &\nodata &0.54 &\nodata &1.08 &\nodata &1.32 &\nodata &K9, L13\\
8 &0 31  $\,$ 1.34 &$-$72 20 30.0 &14.64 &0.28 &0.00 &0.04 &0.80 &0.07 &1.20 &0.12 &HW5 \\
9 &0 32 41.02 &$-$72 34 50.1 &14.66 &0.35 &0.26 &0.18 &0.87 &0.05 &0.90 &0.05 &L14 \\

\enddata

\tablenotetext{a}{K is for \cite{kron}, L is for \cite{lindsay}, HW is for \cite{hw74}, H is for \cite{hodge85}, H86 is for \cite{hodge86}, BS95 is for \cite{bica95}, B is for \cite{brk76}}

\tablecomments{The complete version of this table is at: http://ngala.as.arizona.edu/dennis/rafelski.html  The printed edition contains only a sample.
}
\end{deluxetable*}

Our cluster images are drawn from drift scan imaging of the central
$4^\circ \times 4.5^\circ$ of the SMC in $U, B, 
V$, and $I$ done with the Las Campanas Swope telescope (1m) and the
Great Circle Camera \citep{zsb96} between 1996 November and 1999 December. 
The effective exposure time for any portion of the SMC is between 4 and 5 min 
and the pixel scale is 0.7 arcsec pixel$^{-1}$. 
A photometric catalog of stars is presented by \cite{zar02}, but in this study we use the reduced images rather
than the stellar catalog because of the catalog's incompleteness in the high-density centers of 
stellar clusters. We extract $350^{\prime\prime} \times 350^{\prime\prime}$ subimages
in each of the four filters centered on the cluster from the larger drift scans. We use the photometric solutions that were applied to the stellar catalog,
which have an observational scatter of between 0.01 and 0.04 mag for standard star fields.
This uncertainty is significantly smaller than that resulting from other sources of error in the
measurement of cluster integrated colors.

\subsection{Integrated Cluster Colors}

The measurement of an integrated color or magnitude depends critically on the degree to which
one can subtract the effect of the underlying background. This is one area in
which the availability of moderate resolution digital images should enable a quantitative
improvement over previous studies. The two primary considerations are  where
to set the cluster and background apertures and 
how to remove contamination by stars that are too bright
to be plausible cluster members or random SMC field stars (these 
are typically foreground Galactic stars). 

The optimal choice of aperture size is not evident.
The clusters in this sample vary significantly in size, so a single aperture size
is inappropriate.
Although the cluster's tidal radius might be a natural choice for the outer scale
of the aperture, the measurement of the tidal radius is highly uncertain.
\cite{hill03} found that the radius that encloses 90\% of the light of a cluster, $r_{90}$,
is much less sensitive to uncertainties in defining the background level. We use their
tabulated values of $r_{90}$ in the $V$ band 
to set our aperture size for each cluster independently.
The apertures are circular and are the same for all filter bands.
The background is calculated from the area of the $350^{\prime\prime}
\times 350^{\prime\prime}$ cluster image that lies 5 pixels ($3.5^{\prime\prime}$)
beyond $r_{90}$.

The effect of contamination in both the cluster and background apertures
can be severe. We minimize the impact of the
brightest stars on our calculation of the background
by adopting an upper limit to the pixel values used in the calculation.
First, we calculate the mean background,
without this threshold.  Then we set the threshold to be equal to the mean background level plus
two times the central flux value of the cluster. Lastly, we calculate the mean background level using
this threshold. The possibility of foreground contamination within the cluster aperture is 
smaller because its area is smaller than that of the background aperture. However, from visual 
examination we suspect that the photometry of as many as 73 clusters may be significantly contaminated
by an unusually bright star for at least one of the filters
within the aperture corresponding to the upper $1\sigma$ bounds on $r_{90}$.
Because the measurement of the cluster luminosity is the integrated
flux within the aperture, rather than an average, we cannot simply exclude 
high valued pixels. Instead we
flag such clusters in Table \ref{tab1} but do not  correct their photometry.
We found that these clusters do not have noticeably different
properties than the others.

To arrive at the measurement of the cluster magnitude in each filter, we sum the counts within
the $r_{90}$ cluster aperture, subtract the mean background level, multiply
by 10/9 to correct for the choice of $r_{90}$ as the aperture, and iteratively apply
the photometric solution to incorporate the color terms.
Magnitudes and colors are presented in Table \ref{tab1}.
To estimate uncertainties in these quantities, we recalculate 
the magnitudes and colors using the $1\sigma$ bounds on $r_{90}$.
\cite{hill03} do not present $\sigma _{r_{90}}$
for some clusters because their best fit
King model was not statistically acceptable. For these clusters we 
adopt the average of the color uncertainty calculated for the other clusters. 
The uncertainty from the photometric calibration is added in quadrature,
but it is a minor contribution.
Magnitudes and colors, in particular filter bands, are unavailable for 
the 35 clusters whose images are 
saturated in at least one of the filters, the
two clusters that are off an edge of a scan, and the one cluster (number 180) whose image
quality is too low to be useful.
These problems often affect the cluster image
in only one filter, 
and in such cases the magnitudes are calculated for the unaffected filters.
There are 38 clusters that are missing at least one 
of the three colors in our Table \ref{tab1}, but only seven for which we are unable to compute
any colors. Eliminating these seven plus the two that are off the edges of scans leaves
us with 195 clusters, out of the original 204, for which we measure an age.

Because contamination is such a serious problem, and because, as we will show
below, the integrated colors scatter widely around the model predictions, we
investigate whether adopting a smaller cluster aperture would lead to more robust
colors. Using colors from small apertures assumes that there are
no internal color gradients.
We re-measure colors using an aperture of radius $0.5r_{90}$ and find
no noticeable decline in the scatter of the colors about the model predictions. 
We conclude that the contamination cannot be significantly mitigated by
decreasing the aperture, partially because the scatter is also
due to stochastic effects within the cluster's own population.

\begin{figure}
\plotone{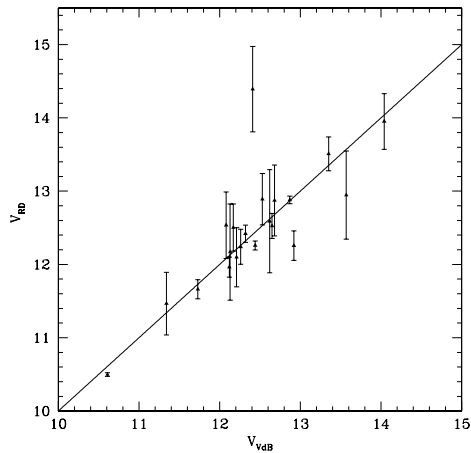}
\caption{Comparison of our V magnitudes, $V_RZ$,  with those from \cite{van81}, $V_VdB$.
The  \cite{van81} compilation did not quote photometric uncertainties, and thus only
uncertainties in our measurements are shown.
The errorbars correspond to those in the tables.
 The line is the 1:1 correlation.
\label{V_Comparison}}
\end{figure}

\begin{figure}
\plotone{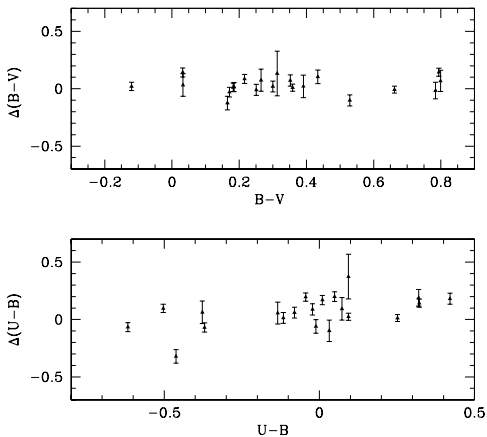}
\caption{Comparison of our $B-V$ and $U-B$ colors with those from \cite{van81}. 
The difference is our values - \cite{van81}.
The  \cite{van81} compilation did not quote photometric uncertainties, and thus only
uncertainties in our measurements are shown.
The errorbars correspond to those in the tables.
\label{color_res}}
\end{figure}

We obtain an external estimate of the measurement uncertainties by comparing our
magnitudes and colors to previous studies where possible. 
A compilation of integrated colors from various
photometric studies is presented by \cite{van81}. 
Figures \ref{V_Comparison} and \ref{color_res} show the differences between our 
V magnitudes and colors and those tabulated by
\cite{van81}. The error bars are underestimated because they
do not include the uncertainties in the compiled data (none were quoted).  
Assuming a standard error of 0.1 mag, the scatter in $B-V$
appears to be entirely consistent
within the uncertainties, while it appears that we have slightly underestimated
the uncertainties in $U-B$ (either in our data or van den Bergh's).
In the mean both colors agree well.
To quantitatively determine whether the scatter is commensurate with the quoted uncertainties,
we examine the distributions of $\Delta$color/$\sigma$. If the uncertainty estimates
are correct, a Gaussian fit to the distribution should have $\sigma = 1$. Instead, we find
(after removing the mean offset, one highly discrepant cluster, and not adopting any
uncertainties for the published data) 
that the fit has a dispersion that is $\sim 1.5$.
However, if the uncertainties in the published data are similar to ours, then $\sigma_\Delta$ 
would be  40\% larger than calculated and the fitted Gaussian
would have a dispersion that is close to one. We conclude that except for
the offset in the mean, our colors are statistically consistent with previous
measurements and our uncertainty estimates are appropriate.

\subsection{Determining Cluster Ages}

The integrated colors of any stellar population provide a luminosity weighted age
measurement. The color should be a particularly stable chronometer for clusters
because they formed  their stars over a timescale that is much shorter
than their age.
We can measure the age of a cluster 
to the degree that we can accurately measure and model the colors.
The colors are presented in Table \ref{tab1}
and the models we use are those of \cite{lei99},  hereafter referred to as the
Starburst99 or S99 models, and of \cite{anders03}, hereafter referred to as the 
GALEV models. 
The S99 models that we use correspond to
the standard mass loss prescription, the
theoretical wind model, and the full isochrone mass interpolation.
We have adopted a Salpeter IMF in both models.
We derive ages using models with 
a range of appropriate metallicites ($Z = $ 0.001, 0.004, and 0.008 for the S99 models and 
0.004 and 0.008 for the GALEV models).
These metallicity values refer to the mass fraction of metals relative to hydrogen and
correspond to [Fe/H] of $-$1.3, $-$0.7, and $-$0.4, respectively. 
Observations of individual clusters
\citep{dopita,dacosta,pagel,def,piatti} and the reconstruction of the 
global chemical enrichment history of the SMC from 
color-magnitude diagrams \citep{hz03} suggest
that the metallicity of the SMC has for the most time varied between [Fe/H] $\sim-1.2$ and $-0.4$.
As we discuss below, some of the finer details of the cluster formation history are extremely
sensitive to the exact choice of metallicity for the model and our current constraints on the
age-metallicity relationship are insufficiently precise to dictate which model
should be used for each age.

\begin{figure}
\plotone{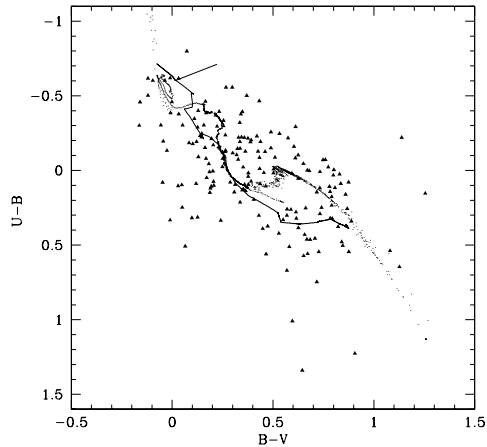}
\caption{Integrated $U-B$ vs $B-V$ from data and models for SMC clusters. 
The triangles represent our measurements, the points, which blend into a line
in certain regions, represent 
the Starburst99 model for Z = 0.004 (for ages between 0.001 and 10 Gyrs), 
and the line represents the GALEV model for Z = 0.004 (for ages between 0.004 and 14 Gyrs). 
\label{color_diagram_ub_bv}}
\end{figure}

\begin{figure}
\plotone{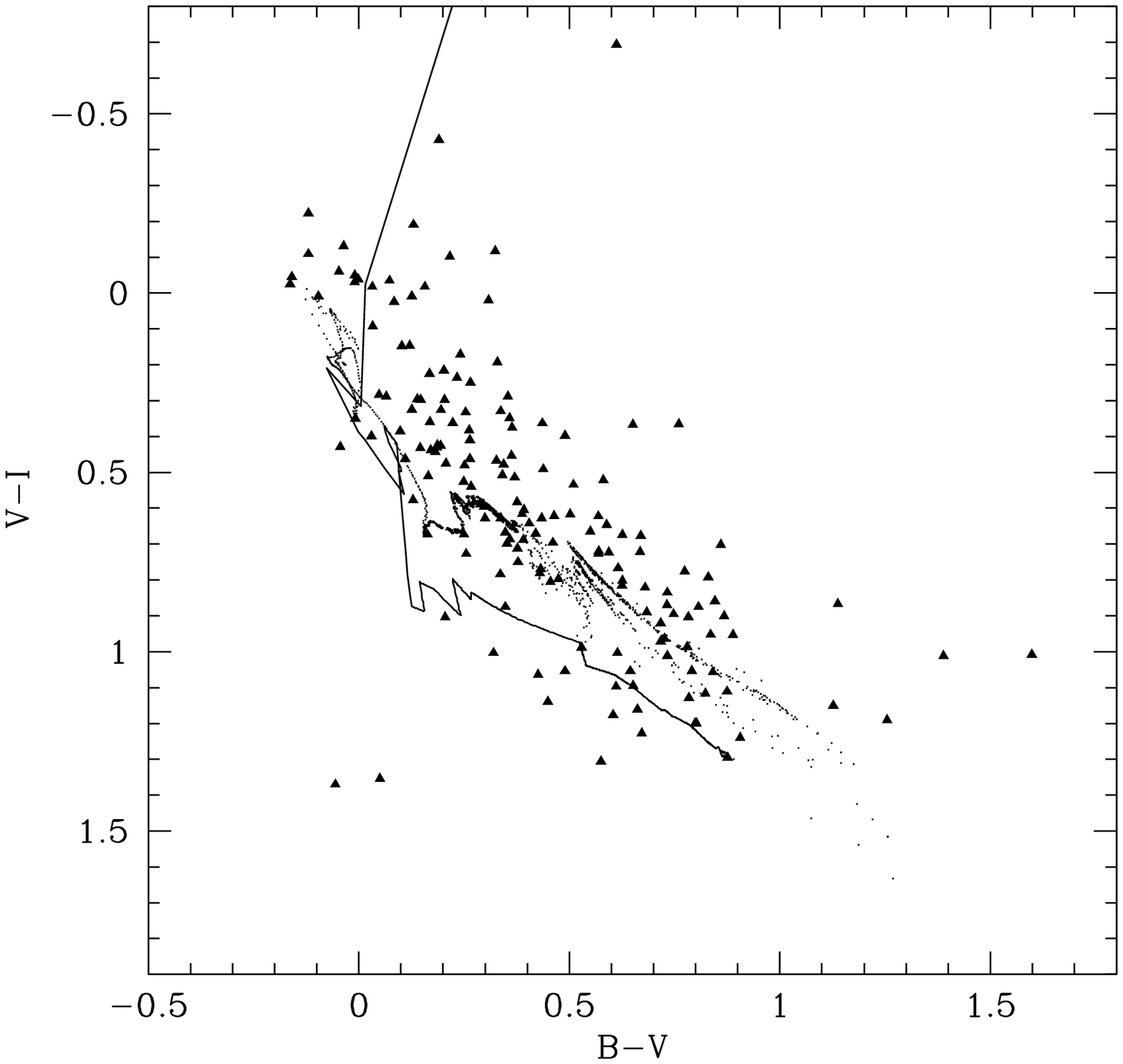}
\caption{Integrated $V-I$ vs $B-V$ from data and models for SMC clusters. 
The triangles represent our measurements, the points, which blend into a line
in certain regions, represent 
the Starburst99 model for Z = 0.004 (for ages between 0.001 and 10 Gyrs), 
and the line represents the GALEV model for Z = 0.004 (for ages between 0.004 and 14). 
\label{color_diagram_vi_bv}}
\end{figure}

\begin{figure}
\plotone{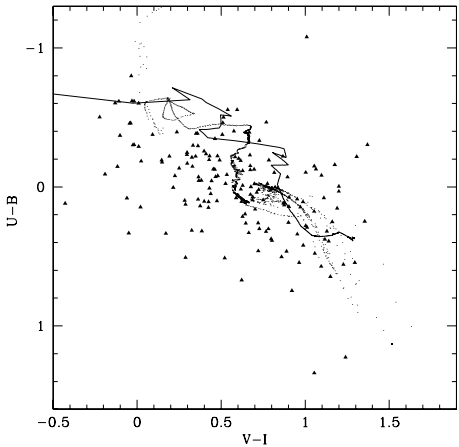}
\caption{Integrated $U-B$ vs $V-I$ from data and models for SMC clusters. 
The triangles represent our measurements, the points, which blend into a line
in certain regions, represent 
the Starburst99 model for Z = 0.004 (for ages between 0.001 and 10 Gyrs), 
and the line represents the GALEV model for Z = 0.004 (for ages between 0.004 and 14). 
\label{color_diagram_ub_vi}}
\end{figure}

In Figures  \ref{color_diagram_ub_bv}, \ref{color_diagram_vi_bv}, and 
\ref{color_diagram_ub_vi} we
overlay the two sets of models over our reddening-uncorrected cluster colors.
We find a mean agreement with the S99 models, although the observational
scatter is large. The principal systematic difference between our data and the models
appears to be in the $V-I$ colors (which are too red in the models for the bluest clusters).
The general agreement of the $B-V$ colors suggests that the problem lies in the modeling
of the $I$-band.  The agreement is significantly poorer relative to the GALEV models, 
but we continue to use those models as well to examine systematic errors introduced
into the derived ages by the use of one or the other set of theoretical models.

\begin{figure}
\plotone{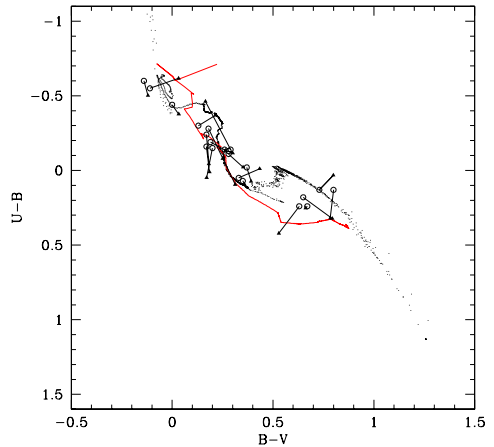}
\caption{Comparison of our color measurements (triangles) to those  
of \cite{van81} (circles). Straight lines connect the measurements for the same cluster. 
The models are as in Figures \ref{color_diagram_ub_bv} - \ref{color_diagram_ub_vi}.
\label{Color_Comparison}}
\end{figure}

The observed scatter of the colors about the theoretical model is the fundamental
observational limit to the precision with which we can measure ages.
The scatter arises from a variety of sources
including the lack of a reddening
correction, which we discuss below, stochastic effects within the cluster population,
contamination of the cluster apertures, 
which we discussed in \S2.2, and the contamination of the background aperture, which
we attempt to minimize.
The last three of these possibilities will become less important as we consider
intrinsically brighter clusters. A comparison (Figure \ref{Color_Comparison})
between our $U-B$, $B-V$ colors and
those presented by \cite{van81}, shows that indeed for these clusters, which are among
the most luminous clusters, the scatter about the model line (in particular the S99
model) is much smaller than for the sample as a whole (Figure \ref{color_diagram_ub_bv}).
We note, however, that the difference between the two models is often comparable to 
the scatter in the data.

To measure a cluster's age,  we calculate the color difference between data, 
typically $U-B$, $B-V$ and $V-I$, although not all colors are available for all clusters
(see \S2.2), and models,
at all ages available for the particular model.  At each age,
$\chi^2$ is evaluated using all the available colors and we adopt the age that corresponds to the minimum $\chi^2$.
Age limits corresponding to the 90\% confidence range are derived using the criterion that $\Delta\chi^2 \le  2.7$.
The best fit ages
and uncertainties are given in Table \ref{tab2},  with the subscript U for reddening-uncorrected
and C for reddening corrected (see below for discussion of the reddening correction).
\begin{deluxetable*}{crrrrrrcr}
\tabletypesize{\scriptsize}
\tablecaption{AGES\label{tbl-2}
\label{tab2}}
\tablewidth{0pt}
\tablehead{
\colhead{Number} &
\colhead{$Age_{U}$}   &
\colhead{$Age_{U}-$} &
\colhead{$Age_{U}+$} &
\colhead{$Age_{C}$}   &
\colhead{$Age_{C}-$} &
\colhead{$Age_{C}+$} &
\colhead{Contamination} &
}

\startdata

1 &1264 &1104 &2352 &1264 &976 &2420 &\nodata \\
2 &\nodata &\nodata &\nodata &\nodata &\nodata &\nodata &\nodata \\
3 &2520 &1368 &8080 &2148 &1160 &11500 &\nodata \\
4 &6280 &5140 &7640 &5940 &3040 &7500 &\nodata \\
5 &2080 &1232 &2920 &1264 &1136 &2220 &X \\
6 &2288 &1264 &4000 &1996 &1148 &3760 &\nodata \\
7 &12020 &7480 &14000 &12020 &5500 &14000 &X \\
8 &924 &724 &1144 &316 &292 &532 &\nodata \\
9 &4060 &2660 &6600 &2740 &1996 &6360 &\nodata \\

\enddata

\tablecomments{Ages are in Myrs. $Age_{U}$ represents uncorrected Ages, and $Age_{C}$ present extinction corrected ages.  An X under contamination denotes a contaminated cluster. The complete version of this table is at http://ngala.as.arizona.edu/dennis/rafelski.html, along with other similar tables for different models and metallicities.  The printed edition contains only a sample.
}
\end{deluxetable*}

Although the age fitting is a maximum-likelihood technique, we also use the minimum
$\chi^2_\nu$ to 
evaluate the goodness of the best-fit model.
There are a 
large number of clusters for which the model fit can be rejected with high
confidence (for example, for the GALEV model with metallicity of 0.004 over half
of the fits could be rejected).
These clusters are not necessarily those with 
poor structural fits or with obviously questionable photometry (for example, they 
are not those that appear to be contaminated by a nearby bright star). Because
the models do not account either for stochastic effects at the top end of the 
stellar luminosity function nor for contamination, we suspect that these poor fits
reflect such problems rather than some inability of the models to reproduce
the cluster stellar populations. Supporting evidence for this conjecture comes
from the similarity in the derived age distribution for clusters with low and high
values of the best-fit $\chi^2_\nu$'s. We include all clusters, regardless of their
minimum $\chi^2_\nu$ value, in subsequent discussion.

One potential cause of high $\chi^2_\nu$ values is the omission, so far, of an extinction correction.
We explore two options for extinction corrections (both adopt a standard Galactic
extinction law \citep{schild} which is acceptable for the SMC at optical wavelengths):
1) we include extinction as a free parameters in our fitting algorithm, and
2) we adopt extinction values based on other data.
Option 1 failed to produce reliable estimates
of the extinction. Because of insufficient observational constraints, 
the algorithm corrected for color scatter in a random way producing both brightening and
dimming effects with larger corrections for more scattered clusters. 
We attribute this failure to that rather small extinction values
expected across the SMC \citep{zar} and to the degree of noise in our color measurements.
Instead, we choose to use the extinction distributions measured by \cite{zar}
using thousands of individual stars across the SMC. The extinction is small 
along most lines of sight, but does vary as a function of stellar type. We adopt
the  approach presented by \cite{hz03} where objects younger than 10 Myr are
assigned the extinction derived from the young stars, objects older than 1 Gyr are assigned
the extinctions derived from the older stars, and objects between these ages are assigned
an extinction that is a linear interpolation (over log age) between the two boundaries.
\cite{hz03} assign objects a random extinction drawn from the observed distribution of
extinctions at each location,
here we simply assign the mean of the distribution. The extinctions are calculated
and added
to the model colors, for which the age, and hence corresponding extinction, is known.
Comparison of $\chi^2$ values proceeds as previously to provide a best-fit measurement
of the age. Although extinction should be included, it does not significantly decrease
the scatter of colors about the models.

\begin{figure}
\plotone{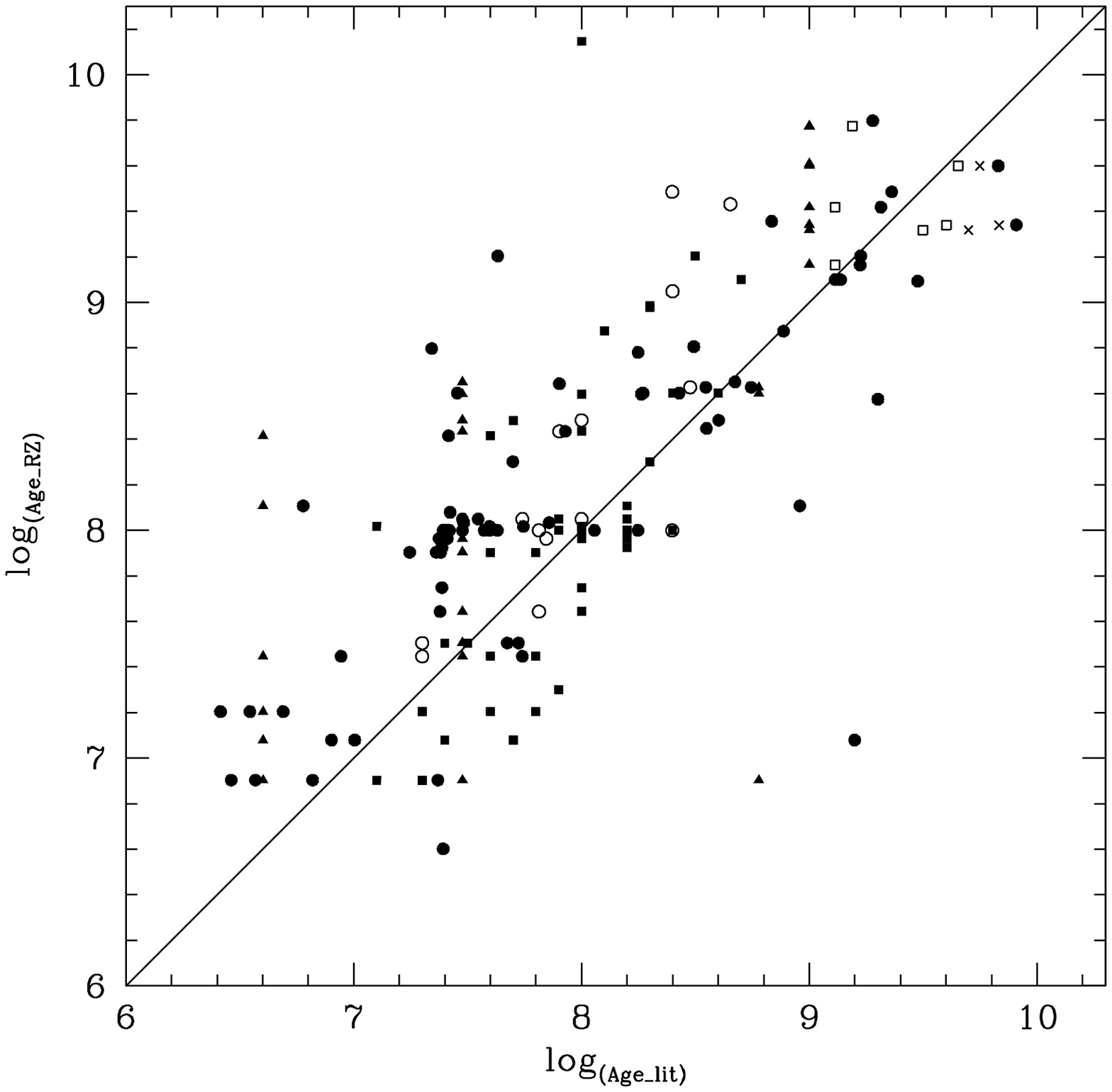}
\caption{Comparison of our measured cluster ages to those available in the literature with our values being Age$_RZ$ and the different literature ages are Age$_lit$.
The various literature studies are \cite{van81}'s (triangles), \cite{pie99}'s (filled boxes), \cite{rich00} (open boxes), \cite{msf98} (crosses), \cite{hunter} (filled circles) and \cite{oliv00}'s (open circles).  
For clarity, error bars are omitted. The line is the 1:1 correlation.
\label{age_comparison}}
\end{figure}

We now compare our derived ages to those presented in the literature. 
The results shown in 
Figure \ref{age_comparison} include
both a comparison to ages obtained via integrated colors \citep{van81, hunter} and 
via isochrone fitting \citep{pie99, oliv00, msf98, rich00}. 
 \cite{van81} provides only an
age range for each cluster and we adopt the midpoint of the range for the comparison. 
For the ages from \cite{msf98} we adopt the primary stated ages, and for the 
 \cite{rich00} ages we used their [Fe/H] of $-$0.71 ages that take into account 
the SMC distance modulus and assume color shifts are due to reddening.
Our ages are broadly correlated to those presented in the literature.
However, when comparing
to specific data sets different patterns emerge. For example, in comparison to the
\cite{van81} data we appear to systematically overpredict the ages. Ignoring any
systematic differences, we calculate that the dispersion about the 1:1 line for all the
samples is 0.76.
This result suggests that the ages are good to a factor of two. However,
comparing only to the most precise ages, those derived from color-magnitude diagrams,
the scatter drops to 0.49, providing added confidence in our measurements.

\section{Discussion}

\subsection{The Cluster Age Distribution for Ages $<$ 5 Gyr}

\begin{figure}
\plotone{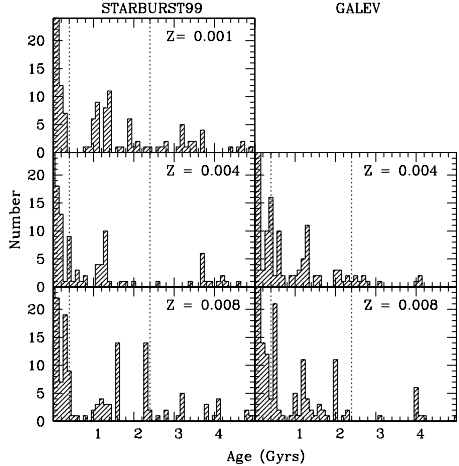}
\caption{Cluster age histograms derived with the Starburst99 and GALEV models.
The bin size is $10^8$ yrs. The dotted lines indicate the ages of measured
field star bursts \citep{hz03}. 
\label{histograms}}
\end{figure}

We present histograms in Figure \ref{histograms} 
of the age distribution of stellar clusters in the SMC.
Depending on the model, there are between 6 and 48 clusters with ages
greater than 5 Gyr that are not shown in this Figure (but are discussed later).  The Figure illustrates two
key points: 1) independent of metallicity and model type, the current age distribution of clusters
is strongly peaked to ages $<$ 1 Gyr (this peak would be further enhanced by the 
inclusion of clusters embedded within emission line regions), 
2) independent of metallicity and model type, the 
distributions are highly variable, although the location of peaks is strongly
sensitive to metallicity and model type.

\begin{figure}
\plotone{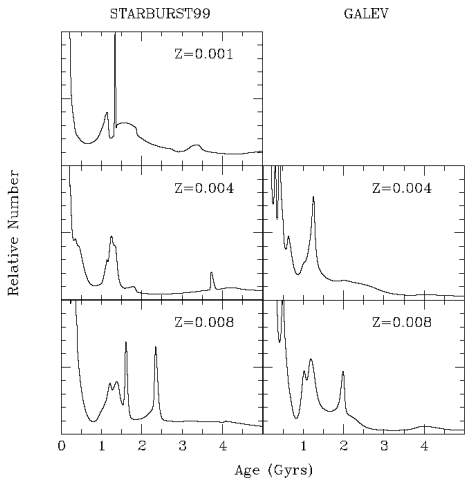}
\caption{Smoothed cluster age distributions derived with the Starburst and GALEV models
(see text for details). 
\label{smooth_histograms}}
\end{figure}

Although a histogram is one way of presenting the data, it fails to convey a sense of
the age measurement uncertainties, which differ greatly among clusters. 
We present the smoothed age distribution in Figure \ref{smooth_histograms},
in which  each cluster is modeled
by a normalized probability distribution that is Gaussian on either side of the best-fit age, 
but in which this asymmetric Gaussian has a dispersion corresponding to the 
appropriate 1$\sigma$ lower or upper value.
One interesting difference between the 
histograms and the smoothed distributions can be seen when comparing
the peaks at $\sim$ 3.5 Gyr in the Z = 0.004 panel and at $\sim$ 3 Gyr in the Z = 0.008 panel for the S99
models. The two peaks are about the same height in the histograms, but very different in the smoothed version. This difference illustrates how the inclusion of uncertainties can
alter the significance of peaks.
The only peaks that appear to reproduce in more than one of the smoothed distributions
are the ones at $\sim$ 0.5 Gyr, at $\sim$1.2 Gyr, and at $\sim$2 Gyr. The dotted lines
in the Figure indicate the position of peaks observed in the global star formation history
of the SMC \citep{hz03}. Interestingly, the 0.5 and 2 Gyr peaks correspond roughly
to the peaks observed in star formation history.
Are any of the peaks seen in the cluster age function statistically significant?

\begin{figure}
\plotone{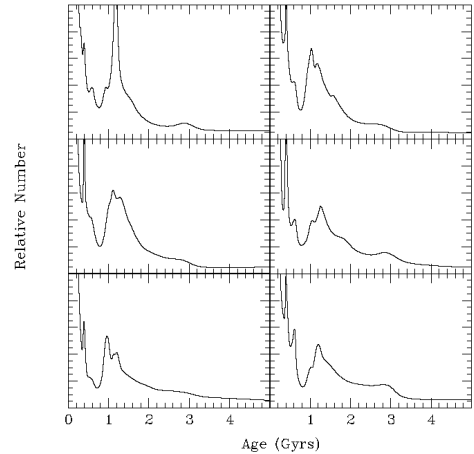}
\caption{Smoothed cluster age distributions derived with the GALEV models from Monte Carlo simulations
(see text for details). 
\label{sim_histograms}}
\end{figure}

The answer to this question is more complicated than simply evaluating the 
number of clusters of a particular age. Our method for estimating ages can
result in structure in the age distribution even if none is present. If the models
traverse a large part of color space quickly and observational errors scatter
clusters within the color space, an unrepresentative large fraction of the clusters
may be close in color to the models that span a small age range. To determine how susceptible 
we are to this problem, we simulate our age measurements assuming
an underlying smoothly exponentially declining cluster age function. 
We make no assumption here as to the cause of the exponential decline (whether
it reflects a true decline in cluster formation or the disruption of clusters). We determine colors
using both the S99 and GALEV models, add uncertainties drawn from our
observational distribution, and
then recover the age distribution using our technique. The resulting distributions
often show some structure, as in the example from the GALEV Z = 0.004 simulations (Figure \ref{sim_histograms}).
To estimate whether the structure in these Monte-Carlo
simulations is as strong as that in the data, we compare
the areas under the peaks.
We automate the measurement of the area and measure peak areas
for both the 
data and 1000 simulations (for each of the model/metallicity combinations). 

Our principal finding is that no peak is consistently significant across the range of
models. While
for certain metallicities, certain peaks are statistically signficant (probability
of random occurrence  $<  0.05$), without
knowing exactly which metallicity model is appropriate we are unable to conclude
that any of the peaks are real.  In short, both
the precision of the colors and the systematic uncertainties inherent in the modeling
preclude any conclusions regarding the fine structure of the cluster age distribution.

\subsection{The Evolution of the Cluster System}

\begin{figure}
\plotone{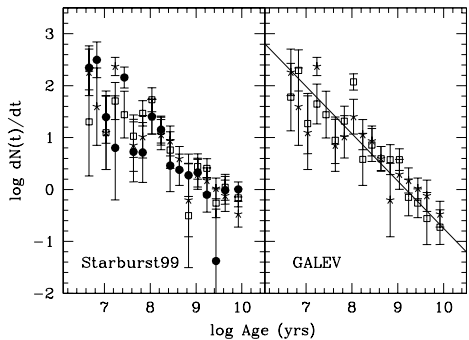}
\caption{The number of clusters of age $t$ per number of stars formed over the equivalent
time. The two panels show results from either Starburst99 (left) or the 
GALEV (right) models. Within each panel the symbols represent the results from
either $Z =$ 0.001 (filled circles), 0.004 (open squares) or 0.008 (crosses). The line
in the left panel represents a power-law distribution with $y \propto t^{-2.1}$.
\label{Fig:dissolve}}
\end{figure}

Under the assumption that the cluster formation history is the same as that measured
for the field stars,
a comparison of the cluster age distribution and the field star formation history
\citep{hz03} provides constraints on the evolution of the cluster system. 
Any variation in the number of clusters normalized by the 
corresponding star formation rate can be interpreted as the result of 
evolution of the cluster system.  Various
factors affect the cluster age distribution, some of which are observational, 
such as the loss from the sample of increasingly fainter or more diffuse clusters with age, 
and some of which are physical, such as the tidal disruption or evaporation of clusters.
We begin by calculating the directly empirical quantity of the number of clusters
of a given age divided by the number of stars formed at that time.  
We bin our clusters into the same age bins as used in the field star formation analysis. 
This calculation provides
a star-formation normalized measure of the differential number of clusters as a
function of time, which approaches $dN(t)/dt$ as the age bin size approaches zero.
If the population of clusters formed
as a fixed fraction of the field star formation and if clusters did not evolve, the ratio
should be constant over time, which is evidently not the case (Figure \ref{Fig:dissolve}).

The number of clusters currently existing as a function of lookback time, 
$dN(t)/dt$, is well-approximated by the power law function, $dN(t)/dt \propto t^{-2.1}$,
shown in the Figure. The best fit slope value for $dN(t)/dt$ as measured using
both the S99 and GALEV models of different metallicities
ranges from $-1.73$ to $-2.18$, although
four of the five slopes cluster between values of  $-2.08$ and $-2.18$. 
In contrast to the details of the cluster age function described in \S3.1, these results
are nearly insensitive to the particular model used.
To provide some intuition for how rapidly the number of clusters in the sample declines with
age we note that  for all the best-fit slopes more than
half the clusters formed are lost  less than 10 Myr after formation, and  
less than 1\% of the clusters survive, where survive is simply defined as being
part of the current sample, past an age of 1 Gyr.

The observed decline in the number of older clusters is due to a combination
of selection and physical effects. Due to our selection of clusters as resolved
stellar systems in which we observe stars well down the luminosity function, we do not
expect strong age-dependent selection effects unless the structural properties
of the clusters, such as the central surface density, evolve strongly with time.
Our conclusion is similar to that reached by \cite{bl} for the \cite{hodge87}
SMC cluster sample. If this claim is correct, then the principal cause of the decline
in the number of clusters is physical. 

This interpretation is supported by the 
general agreement we find with previous studies of cluster disruption both in the SMC \citep{hunter,bl} and other
galaxies \citep{bl}. To be specific, \cite{bl}
have measured cluster disruption in a set of galaxies, including the SMC. In their
modeling of the problem, they find that they can describe the slope of the cluster
age function to be $(1-\alpha)/\gamma$, where $\alpha$ is the slope of the cluster
initial mass function and $\gamma$ describes the power-law mass dependence of the 
disruption time, $t_d({M}) \propto M^\gamma$. They find that on average $\gamma = 
0.57 \pm 0.1$. From our slope of $-2.1$ in the SMC, we calculate that $\gamma = 0.48$
for an assumed $\alpha = 2$, which is the value of $\alpha$ also assumed by \cite{bl}. 
Alternatively, we could adopt the mean value of $\gamma$ presented by \cite{bl} and calculate
$\alpha = 2.2$, in agreement with the typical cluster initial mass function found
in other galaxies \citep{zf,whit,bik03} and within the range of 2 to 2.4 found for the
LMC and SMC by \cite{hunter}.

One aspect in which we do identify a disagreement with previous studies is in the time at which 
disruption dominates over fading for the SMC cluster population. \cite{bl}, using data from
\cite{hodge87}, identify a bend in $dN(t)/dt$, at $\log{t} = 8.8$, that is associated with this
transition. We find no unambiguous flattening of $dN(t)/dt$ at that, or any other, time. 
However, due to uncertainties, we cannot rule out
a flattening at $\log{(t)} < 8.5$ for the S99 models, or   
for $\log(t) < 7$ in any of our models. The latter could correspond to  
a disruption timescale that agrees with
what \cite{bl} find for other galaxies.  As discussed in \S2.1, we have a strong
bias {\sl against} finding young clusters, and therefore the $dN(t)/dt$ function
is likely to be even steeper at young ages than what we have plotted in
Figure \ref{Fig:dissolve}. The lack of a strong flattening signature and the uncertainty posed
by incompleteness preclude us from placing strong constraints on any flattening
at $\log(t) < 8.5$.

\subsection{Investigating the Age Gap}

The field star formation history \citep{hz03} shows two pronounced peaks at 
about 0.4 and 2.5 Gyr. As we described above, such peaks are seen in the
cluster age distribution {\sl for certain choices of model type and metallicity}. While
we cannot confirm that these peaks exist in the cluster age function, neither can we
exclude the possibility. Our inability to resolve peaks is a consequence of the
low precision of our age estimates. However, another interesting feature of
cluster age functions, particularly in the LMC, is the presence of an ``age gap".
A similar lull in star formation is observed in the global star formation of 
the SMC \citep{hz03}. Can we reach any conclusion about cluster formation over
the timescale of this lull (from 3 to 8 Gyr)?

\begin{figure}
\plotone{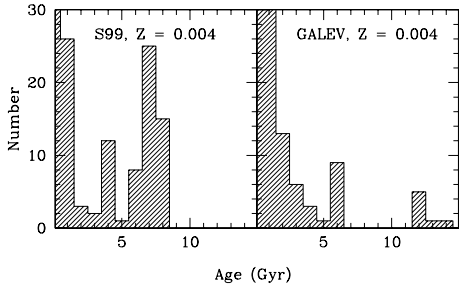}
\caption{The number of clusters of age $t$.
The two panels include the results from either Starburst99 (left) or the 
GALEV models (right) for Z = 0.004.
\label{gap1}}
\end{figure}

In Figure \ref{gap1} we show the distribution for cluster ages between 1 and 15 Gyr
(there are no clusters at measured ages $>$ 15 Gyr) for both the S99 and
GALEV models with Z = 0.004. The distributions qualitatively show an initial,
older epoch of clusters and a recent epoch of cluster formation (starting within 3 Gyr of the 
current time), with a relatively
quiet time between. The quantitative details, however, are quite different between
the models.
First, the oldest clusters in the S99 models are only 8 Gyr old. If this is correct, 
or if we are incomplete in older clusters, then this peak would correspond to 
the $\sim$ 8 Gyr clusters  identified by \cite{rich00}. On the other hand, 
if there is a systematic error in the entire scale of ages, and the older peak corresponds to the
oldest LMC clusters  which are $>$ 10 Gyr old, then perhaps the intermediate
peak (currently located at $\sim$ 4 Gyr) corresponds to the 8 Gyr clusters.  Curiously, the 
GALEV models also show an intermediate age peak (at $\sim$ 6 Gyr), which could correspond to 
the hypothesized 8 Gyr population, and these models do identify a truly old, $>$ 10 Gyr,
population of SMC clusters. In both sets of models, the SMC appears to have had
an initial cluster formation epoch followed by relatively little cluster formation, with
the possible exception of an intermediate age peak, punctuated  with a recent
($<  3$ Gyr old) episode of cluster formation that is at least as strong as the original
epoch.

\begin{figure}
\plotone{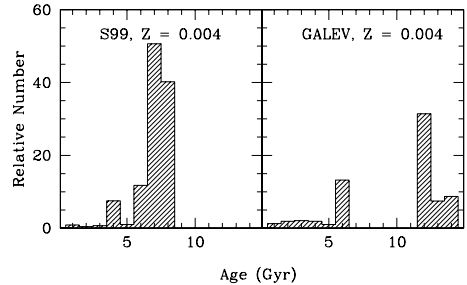}
\caption{The number of clusters of age $t$ normalized by the 
expected dissolution function, $\Delta N(t) \propto t^{-2.1}$.
The two panels include the results from either Starburst99 (left) or the 
GALEV models (right) for Z = 0.004. 
\label{gap2}}
\end{figure}

A different picture of the cluster history is presented if we remove the effect
of cluster dissolution. In Figure \ref{gap2} we have presented the entire cluster
age distribution, but we have normalized by the $t^{-2.1}$ dependence of the 
cluster number. From this Figure it is evident that the initial episode of cluster
formation dominates the rate of cluster formation in the SMC. This Figure rests
on the assumption that the cluster destruction function remains constant in time.
From either model, one can conclude that the initial flurry of cluster formation
(at 11 to 14 Gyr in the Galev models or 6-8 Gyr in the S99 models) did subside,
but whether there is an age gap (as seen for  6 to 11 Gyr in the Galev models)
or rather a continuous low level of cluster formation (as seen for $t \le 5$ Gyr in the
S99 models) depends on the model.
The more dramatic impression of 
an age gap presented in Figure \ref{gap1} is due to the fact that the clusters
formed in the most recent episode of
cluster formation have had insufficient time to dissolve.

\subsection{Spatial Distribution}

\begin{figure}
\plotone{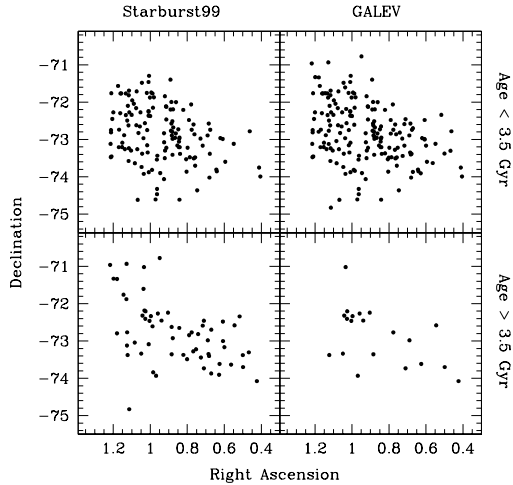}
\caption{The distribution of young (age $<$ 3.5 Gyr) and old (age $>$ 3.5 Gyr) clusters
using ages determined with either the S99 or GALEV models (with Z = 0.004).
\label{spatial}}
\end{figure}

The distribution of clusters of different ages may constrain models of how clusters
formed or were destroyed. \cite{van91} shows that among his sample of SMC clusters
the young clusters lie along the SMC ``bar", while the older clusters form more of 
halo. Although the dynamical reality of the bar has been questioned \citep{zar00}, the distribution
of younger and older stars in the SMC follows a similar pattern in that the younger
stars are preferentially  found along the north-east/south-west axis while the older
stars lie in a spheroidal distribution. In Figure \ref{spatial} we plot the
distribution of clusters divided into a young (age $<$ 3.5 Gyr) and old (age $>$ 3.5 Gyr)
samples\footnote{This division is quite different than that adopted by \cite{van91}.
He chose a color cut that for our models corresponds to $5.1 \times 10^8$ and
$6.6\times 10^8$ for the S99 and GALEV models respectively. Therefore, his conclusions
should not be compared to our Figure \ref{spatial}.}
for S99 and GALEV models with Z = 0.004. 
The characteristics of the distributions, centroid, ellipticity, and position angle are
different, particularly for the S99 model, which has more clusters in the older
bin. For example, the right ascension centroids of the young and old population are
highly discrepant ($0.942 \pm 0.015$ and $0.860 \pm 0.028$ in decimal hours, respectively, which 
corresponds to a difference that is different than zero at more than 2.6$\sigma$ significance),
while  the declination centroids are consistent ($-72.853 \pm 0.07$ and$-72.791 \pm 0.11$ in 
decimal degrees).

Because of the poor temporal resolution of our age measurements, we are
unable to confidently explore other age bins, but such a study with higher
precision ages may help resolve the origin of the gas that is hypothesized
to have infallen \citep{zh03} and the disturbed morphology of the young stars
\citep{zar00}. Correlations among spatial properties and marginally detected age
peaks would also increase the
confidence of any identified structure in the cluster age function.

\section{Conclusions}

We use the images from the Magellanic Clouds Photometric Survey \citep{zar02} to
measure integrated $U, B, V$ and $I$ colors for a sample of 195 stellar clusters
with measured structural parameters \citep{hill03}. We estimate cluster ages
for 195 of these using those colors and models of simple stellar populations. We conclude that:

1) Although peaks are visible in the cluster age distribution, the location of these
varies sensitively with the choice of metallicity and model type, and similar peaks can
arise by chance.

2) We find that the rate of cluster number evolution for clusters of age $t$, after normalizing by the 
number of stars formed at age $t$, is given by $dN(t)/dt \propto t^{-2.1}$ .

3) If this decline in the number of clusters with age
 is attributed primarily to physical destruction of the clusters, we
calculate that the disruption timescale at a fixed mass is $t_d(M) \propto M^{0.48}$ for
a cluster initial mass function $N(M) \propto M^{-2}$.  This result is in agreement with
the average value measured in a set of nearby galaxies, $M^{0.57\pm0.1}$ \citep{bl}.
Alternatively, when we adopt the average value for the disruption timescale from \cite{bl},
we calculate that the initial mass function 
$N(M) \propto M^{-2.2}$. These values are consistent
with results from previous studies on other galaxies and the SMC. The agreement implies that
cluster dissolution is highly effective and relatively independent of the host galaxy.
Only 1\% of clusters formed in the SMC will survive beyond 1 Gyr. This makes comparison
of properties between clusters in the SMC and other systems, like the Milky Way, very difficult unless
ages are known for all the clusters.

4) There is no evidence of a significant age gap in the SMC, although the dominance of the 
initial epoch of cluster formation in conjunction with the rapid dissolution of 
clusters produces an observed cluster age function that appears to have a formation lull
at intermediate ages.

5) The spatial distribution of the younger and older clusters is statistically different,
and supports the hypothesis of a significant accretion or merger event at around
3 to 5 Gyr.

The principal shortcoming of this study is the lack of precision in our age measurements
that are both the result of photometric integrated colors that are susceptible to large
scatter, some of which may be inherent to the cluster population, and cluster models
that do not predict consistent colors for a given age, partly due to the lack of metallicity
constraints. This problem of age and metallicity degeneracy was also examined by \cite{dg03} in 
their study of the systematic uncertainties in cluster age determinations 
in more distant galaxies using aperture photometry and stellar population models.
The Magellanic Cloud clusters are particularly valuable in addressing these issues
because ages from stellar color-magnitude diagrams can be compared to those
of integrated colors (\S2.3). The current limitation is the small number of 
cluster color-magnitude diagrams currently available.
To further address the question of systematics, our study can provide guidance 
as to which clusters to target for more precise age measurements. Such observations
benefit both the study of the Magellanic Clouds and of more distant systems.

\section{Acknowledgments}
We thank the anonymous referee for a careful reading of the manuscript.
MR acknowledges financial support from the Baird Foundation and scholarships from the University of Arizona. 
DZ acknowledges financial support from National Science Foundation
CAREER grant AST-9733111, AST-0307482, and a fellowship from the David and Lucile
Packard Foundation.

\clearpage

\end{document}